\documentclass[reprint,amsmath,amssymb,aps,superscriptaddress,nofootinbib]{revtex4-1}
\usepackage{graphicx}
\usepackage{dcolumn}
\usepackage{bm}
\usepackage[utf8]{inputenc}
\usepackage{float}
\usepackage{hyperref}
\renewcommand{\equationautorefname}{Eq.}

\usepackage{amsmath, amsthm, amsfonts,amssymb}
\usepackage[svgnames]{xcolor}
\usepackage{mathtools}
\usepackage{caption}
\usepackage{subcaption}
\usepackage{xcolor}
\usepackage{soul}
\usepackage{xcolor}
\usepackage{appendix}
\usepackage{orcidlink}
\usepackage{booktabs}

\definecolor{teal}{RGB}{0, 128, 128}
\hypersetup{ colorlinks=true, linkcolor=teal, citecolor=teal, urlcolor=teal}
\makeatletter
\AtBeginDocument{%
  \let\plain@equationautorefname\equationautorefname
  \def\equationautorefname{\plain@equationautorefname\@autoref@insert@tagform}%
  \def\@autoref@insert@tagform~#1\null{~(#1)\null}%
}
\makeatother

\begin{document}

\title{Scalar Bosons with Coulomb Potentials in a Space with Dual Topological Defects in Rainbow Gravity}

\author{L. G. Barbosa \orcidlink{0009-0007-3468-3718}}
\email{leonardo.barbosa@posgrad.ufsc.br}
\affiliation{Departamento de Física, CFM - Universidade Federal de \\ Santa Catarina; C.P. 476, CEP 88.040-900, Florianópolis, SC, Brazil}

\author{J. V. Zamperlini \orcidlink{0009-0002-9702-1555}}
\email{joao.zamperlini@posgrad.ufsc.br}
\affiliation{Departamento de Física, CFM - Universidade Federal de \\ Santa Catarina; C.P. 476, CEP 88.040-900, Florianópolis, SC, Brazil}

\author{L. C. N. Santos \orcidlink{0000-0002-6129-1820}}
\email{luis.santos@ufsc.br}
\affiliation{Departamento de Física, CFM - Universidade Federal de \\ Santa Catarina; C.P. 476, CEP 88.040-900, Florianópolis, SC, Brazil}

\begin{abstract}
This work studies the relativistic quantum dynamics of scalar bosons in a spacetime containing both a cosmic string and a global monopole within the framework of Rainbow Gravity. An effective metric is constructed to describe the combined topological defects together with the energy-dependent deformation of spacetime. The Klein–Gordon equation is formulated in this background, including scalar, vector, and nonminimal couplings, and its solutions are obtained by separation of variables. Generalized Coulomb-type interactions are considered, allowing a unified analysis of scattering and bound states. The bound-state spectrum is determined from the poles of the corresponding $S$-matrix. Two specific choices of rainbow functions are examined, and their influence on the energy spectrum is analyzed through numerical calculations and, in suitable limits, analytical approximations. The results show how the interplay between topological defects and rainbow gravity corrections affects the spectral properties of scalar bosons, while known results are consistently recovered in appropriate limits.
\end{abstract}

\maketitle

\section{Introduction}\label{Introduction}

Topological defects such as cosmic strings and global monopoles are natural outcomes of symmetry-breaking phase transitions in the early universe \cite{string12,string13,string14,string15,string16}. Cosmic strings are linear defects characterized by a mass per unit length, whose gravitational field generates a spacetime that is locally flat but globally conical. This peculiar geometry leads to observable effects such as gravitational lensing and the gravitational Aharonov–Bohm effect. Global monopoles, on the other hand, are pointlike defects associated with a nontrivial vacuum configuration, giving rise to a spacetime with a deficit solid angle. Both defects alter the structure of spacetime and, consequently, influence the propagation and dynamics of quantum fields.

The impact of cosmic string backgrounds on quantum systems has been widely explored in the literature. Studies involving the Schr\"odinger equation \cite{wang2015exact,muniz2014landau,ikot2016solutions,ahmed2023effects}, the Klein–Gordon equation \cite{santos2,string10,string9,string6,string5,string4,string2,neto2020scalar,boumali2014klein,ahmed2021effects,santos2018relativistic}, and the Dirac equation \cite{inercial10,inercial8,string8,string7,string1,hosseinpour2015scattering,marques2005exact,bakke2018dirac,hosseinpour2017scattering,cunha2020dirac,lima20192d} have shown that cosmic strings can modify energy spectra, lift degeneracies, and affect scattering processes. In a similar spirit, global monopole spacetimes have been shown to influence the behavior of non-relativistic particles \cite{global4,ahmed2024effects,alves2023approximate,alves2024exact,mustafa2023schrodinger,BezerradeMello:1996si}, relativistic bosons \cite{global2,de2006exact,de2022klein,montigny2021exact,bragancca2020relativistic,ahmed2022relativistic}, and fermionic fields \cite{global3,ali2022vacuum,ren1994fermions,bezerra2001physics}.

More generally, the study of quantum systems in curved spacetimes, including the role of noninertial effects \cite{inercial1,inercial2,inercial4,inercial5,inercial6,inercial7,inercial9,inercial10,inercial11,santos3,santos2019klein,santos2023non}, has clarified how spacetime geometry modifies fundamental quantum equations. A representative example is the hydrogen atom on curved backgrounds, where curvature-induced corrections to the energy spectrum arise when compared to flat-spacetime results \cite{parker1}. Such corrections stem from geometric effects rather than standard gravitational redshift or Doppler shifts. Related investigations have addressed neutrino propagation in curved spacetimes \cite{neutrino1}, particle emission in backgrounds containing topological defects \cite{jusufi2015scalar}, and scalar bosons interacting through Coulomb-type potentials in cosmic string geometries, where relevant modifications to bound states and scattering amplitudes were reported \cite{neto2020scalar}.

Recent works have further considered the Klein–Gordon equation in G{\"o}del-type spacetimes pierced by a cosmic string \cite{string11}, showing that the defect removes degeneracies in the energy spectrum for Som–Raychaudhuri, spherical, and hyperbolic geometries. The influence of external magnetic fields in related settings has also been investigated \cite{santos1}. These studies reinforce the role of topological defects as key elements in shaping quantum dynamics.

Despite the extensive literature, most analyses focus on a single topological defect at a time. The combined effects of different defects, such as the simultaneous presence of a cosmic string and a global monopole, have received comparatively little attention. In this work, we address this issue by studying the relativistic quantum dynamics of scalar bosons in a spacetime containing both defects. We analyze the Klein–Gordon equation with generalized Coulomb interactions in this mixed background, emphasizing how the combined geometry modifies the equations of motion, phase shifts, and the corresponding $S$-matrix. This framework generalizes previous results obtained for isolated defects and allows known solutions to be recovered as particular limits.

Scalar and vector interactions are introduced through standard couplings, and an additional non-minimal interaction is included to extend the structure of the model. Bound states are identified from the poles of the $S$-matrix, providing a clear connection between the energy spectrum and the parameters associated with each topological defect.

The effects of quantum gravity can be introduced in the context of Rainbow gravity, where the spacetime metric depends on the energy of the probe particle. This dependence leads to modified dispersion relations and, consequently, alters the energy spectra of quantum systems \cite{santos2025scalar}. The resulting changes are especially relevant in the analysis of wave equations, whose solutions are sensitive to the specific choice of rainbow functions. Studies in this context show that non-trivial geometries and topological backgrounds can significantly influence the spectral properties of fields when quantum gravity corrections are considered. In this context, an extension of the studies of scalar fields involving topological defects to rainbow gravity proves quite interesting, which is precisely the objective of this work, in which we extend the results of \cite{Barbosa:2025hva} considering rainbow gravity.

The paper is organized as follows: \autoref{Double_defect_spacetime_in_gravity’s_rainbow} presents the cosmic string and global monopole spacetimes and constructs the effective metric describing their combined geometry. In \autoref{Klein-Gordon_equation_in_the_context_of gravity's_rainbow}, the Klein–Gordon equation is derived in this background, and its general solutions are discussed. \autoref{Generalized_Coulomb_Potentials} analyzes generalized Coulomb potentials and their effects on bound and scattering states. Specific choices of scalar and vector couplings, together with the conditions for bound states, are examined in \autoref{Scattering_and_bound_states}. Secs.~\ref{Case_I} and~\ref{Case_II} are devoted to analytically and numerically studying relevant choices of rainbow functions. Finally, \autoref{Discussion_and_conclusions} summarizes the main results.

\section{Double defect spacetime in gravity’s rainbow}\label{Double_defect_spacetime_in_gravity’s_rainbow}
Topological defects, such as cosmic strings and global monopoles, can influence spacetime geometry. When both defects coexist, their combined effects can lead to new geometric features and alter the behavior of physical systems. We consider a generalized spacetime metric that incorporates both defects through their deficit angles.

The metric for a spacetime with a cosmic string and a global monopole, in spherical coordinates, is given by\footnote{Throughout the paper, Planck units $(G=c=\hbar=1)$ are used, so that all quantities are dimensionless.} \cite{bezerra2002bremsstrahlung,jusufi2015scalar}:
\begin{equation}
    ds^{2} = -dt^{2} + dr^{2} + \beta^{2} r^{2} d\theta^{2} + \alpha^{2} \beta^{2} r^{2} \sin^{2}\theta d\phi^{2},
\end{equation}
where $\alpha = 1 - 4G\mu$ and $\beta^{2} = 1 - 8\pi G\eta^{2}$, with $\mu$ and $\eta$ representing the mass density and symmetry-breaking scale, respectively. We explore the effects of these defects within gravity's rainbow \cite{Momeni:2017cvl}, a modification of general relativity that accounts for energy-dependent spacetime geometry. The modified line element is given by
\begin{equation}
   ds^{2} = -\frac{dt^{2}}{f^{2}(x)} + \frac{dX^{2}}{g^{2}(x)},
\end{equation}
where $x = \xi \varepsilon / \varepsilon_p$, and $f(x)$, $g(x)$ are rainbow functions satisfying $f(0) = g(0) = 1$ and $\xi$ a constant parameter. In the presence of topological defects, the line element becomes
%aqui
\begin{equation}
    ds^{2}=-\frac{dt^{2}}{f^{2}\left(x\right)}+\frac{dr^{2}+\beta^{2}r^{2}d\theta^{2}+\alpha^{2}\beta^{2}r^{2}\sin^{2}\left(\theta\right)d\phi^{2}}{g^{2}\left(x\right)}
\end{equation}
which incorporates information about the topological defects through the parameters \(\alpha\) and \(\beta\), as well as the energy-dependent effects of gravity's rainbow via the rainbow functions $f(x)$ and $g(x)$.

\section{Klein-Gordon equation in the context of gravity's rainbow}\label{Klein-Gordon_equation_in_the_context_of gravity's_rainbow}
We consider the covariant Klein--Gordon equation for a scalar boson of mass $M$, in the presence of a scalar coupling $V_s$, a vector potential $A_\mu$, and an oscillator-like interaction $X_\mu$, which can be written as
\begin{equation}\label{KG}
-\frac{1}{\sqrt{-g}}\, D_{\mu}^{(+)}\!\left( g^{\mu\nu}\sqrt{-g}\, D_{\nu}^{(-)} \psi \right)
+ \left( M + V_s \right)^2 \psi = 0,
\end{equation}
where the differential operator is defined as $D_{\mu}^{(\pm)} = \partial_{\mu} \pm X_{\mu} - i e A_{\mu}$. Now, we consider the specific case where $A_{\mu} = (A_t(r), 0, 0, 0)$ and $X_{\mu} = (X_t(r), X_r(r), 0, 0) $. By substituting the metric coefficients and determinant, the Klein-Gordon \autoref{KG} becomes
\begin{equation}
\begin{aligned}
&-\frac{f^{2}}{g^{2}}\left(\frac{\partial^{2}}{\partial t^{2}}-2ieA_{t}\frac{\partial}{\partial t}-e^{2}A_{t}^{2}-X_{t}^{2}\right)\psi \\
&\quad +\frac{\partial^{2}\psi}{\partial r^{2}}+\frac{2}{r}\frac{\partial\psi}{\partial r}-\left(\frac{\partial X_{r}}{\partial r}+\frac{2}{r}X_{r}+X_{r}^{2}\right)\psi\\
&\quad +\frac{1}{\beta^{2}r^{2}}\frac{1}{\sin\left(\theta\right)}\frac{\partial}{\partial\theta}\left(\sin\left(\theta\right)\frac{\partial\psi}{\partial\theta}\right) \\
&\quad +\frac{1}{\beta^{2}r^{2}}\frac{1}{\alpha^{2}\sin^{2}\left(\theta\right)}\frac{\partial^{2}\psi}{\partial\phi^{2}}-\frac{\left(M+V_{s}\right)^{2}}{g^{2}}\psi=0.
\end{aligned}
\end{equation}

Considering the symmetry of the problem, we separate variables and adopt the following ansatz for the wave function:
\begin{equation}
   \psi(t,r,\theta,\phi) = \frac{u(r)}{r} S(\theta) e^{im\phi} e^{-i\varepsilon t},
\end{equation}
where \( m = 0, \pm1, \pm2, \pm3, \dots \) and \(\varepsilon\) denotes the energy of the scalar boson. This choice isolates the radial and angular dependence and makes manifest the azimuthal symmetry associated with the Killing vector \(\partial_{\phi}\).

Substituting the ansatz into the wave equation yields the angular equation
\begin{multline}
    \frac{1}{\sin(\theta)} \frac{d}{d\theta} \left( \sin(\theta) \frac{dS(\theta)}{d\theta} \right) \\
    + \left( l_{\alpha} (l_{\alpha}+1) - \frac{m^2}{\alpha^2 \sin^2(\theta)} \right) S(\theta) = 0,
\end{multline}
which is the standard polar equation with a modified azimuthal term. Its solutions are written in terms of generalized Legendre functions \(P^{\mu}_{\nu}(x)\); the relevant parameter relations are
$l_{\alpha} = l + |m|\left( \frac{1}{\alpha} - 1 \right)$, $l = n + |m|$, $m_{\alpha} = \frac{m}{\alpha}$ and $-l_{\alpha} \leq m_{\alpha} \leq l_{\alpha}$.

These relations encode the shift in the effective angular momentum induced by the parameter $\alpha$ and determine the allowed angular quantum numbers. 

Turning to the radial sector, one obtains a Schrödinger-like equation for $u(r)$,
\begin{equation}
    \frac{d^2 u(r)}{dr^2} + \left( K^2 - V_{\text{eff}}^2 - \frac{l_{\alpha}(l_{\alpha}+1)}{\beta^2 r^2} \right) u(r) = 0,
\end{equation}
where the auxiliary quantity $K^{2}$ is defined as follows
\begin{equation}
    K^{2}=\frac{f^{2}\varepsilon^{2}-M^{2}}{g^{2}},
\end{equation}
and $V_{\text{eff}}^{2}$ defines the effective potential,
\begin{multline}
    V_{\text{eff}}^{2}=\frac{V_{s}^{2}-f^{2}e^{2}A_{t}^{2}+2\left(MV_{s}-f^{2}e\varepsilon A_{t}\right)}{g^{2}} \\
    +\frac{dX_{r}}{dr}+\frac{2}{r}X_{r}+X_{r}^{2}-\frac{f^{2}}{g^{2}}X_{t}^{2}.
\end{multline}

Thus, the radial problem reduces to finding eigenfunctions $u(r)$. We emphasize that $A_{t}$, $V_{s}$, $X_{t}$, and $X_{r}$ are arbitrary functions and therefore enter the radial dynamics only through the combination shown above.

\section{Generalized Coulomb Potentials}\label{Generalized_Coulomb_Potentials}
We analyze the Coulomb-type interaction for the generalized potentials. In this context, considering a class of generalized Coulomb potentials, we define  
\begin{equation}
    A_{t} = \frac{\gamma_{t}}{r}, \quad V_{s} = \frac{\gamma_{s}}{r}, \quad X_{t} = \frac{\delta_{t}}{r}, \quad X_{r} = \frac{\delta_{r}}{r},
\end{equation}
where $\gamma_t$, $ \gamma_s$, $\delta_t$, and $\delta_r$ denote coupling parameters. Consequently, the radial equation governing the dynamics takes the form  
\begin{equation}
   \frac{d^{2}u\left(r\right)}{dr^{2}}+\left(K^{2}-\frac{\alpha_{1}}{r}-\frac{\alpha_{2}+l_{\alpha}\left(l_{\alpha}+1\right)/\beta^{2}}{r^{2}}\right)u\left(r\right)=0,
\end{equation}
where we introduce the following definitions:
\begin{eqnarray}
\alpha_{1} &=& 2\left(\frac{M\gamma_{s}-f^{2}e\varepsilon\gamma_{t}}{g^{2}}\right), \\
\alpha_{2} &=& \delta_{r}\left(\delta_{r}+1\right)
+ \frac{\gamma_{s}^{2}}{g^{2}}
- \frac{f^{2}}{g^{2}} e^{2}\gamma_{t}^{2}
- \frac{f^{2}}{g^{2}} \delta_{t}^{2}.
\end{eqnarray}

To solve the radial equation, we perform the change of variable \( z = -2iKr \), mapping it to a Whittaker-type equation:
\begin{equation}
    \frac{d^{2}u(z)}{dz^{2}} + \left(-\frac{1}{4} - \frac{i\eta}{z} + \frac{\frac{1}{4} - \gamma_{l}^{2}}{z^{2}}\right) u(z) = 0,
\end{equation}
where we define
\begin{eqnarray}
\gamma_{l}^{2} &=& \frac{1}{4}
+ \frac{l_{\alpha}(l_{\alpha}+1)}{\beta^{2}}
+ \alpha_{2}, \\
\eta &=& \frac{\alpha_{1}}{2K}.
\end{eqnarray}

The general solution of the radial equation can be written as a linear combination of Whittaker functions,
\begin{equation}
    u(z) = A M_{-i\eta,\gamma_{l}}(z) + B W_{-i\eta,\gamma_{l}}(z),
\end{equation}
where the two independent Whittaker branches correspond to different asymptotic behaviours at the origin and at infinity. Requiring regularity at the origin excludes the $W$-branch and thus fixes $B=0$. Consequently, the physical radial solution reduces to
\begin{equation}
    u(z) = A e^{-\frac{z}{2}} z^{\frac{1}{2} + \gamma_{l}} M\left( \frac{1}{2} + \gamma_{l} + i\eta, 1 + 2\gamma_{l}, z \right),
\end{equation}
where $M(a,b;z)$ denotes Kummer’s confluent hypergeometric function of the first kind, which completely determines the radial behaviour.

\section{Scattering and bound states}\label{Scattering_and_bound_states}

For large $|z|$ the confluent hypergeometric function $M(a,b;z)$ has the asymptotic form
\begin{multline}
     M(a, b; z) \simeq \frac{\Gamma(b)}{\Gamma(b - a)} e^{-\frac{i}{2} \pi a} | \overline{z} |^{-a}
    \\+ \frac{\Gamma(b)}{\Gamma(a)} e^{-i\left( | \overline{z} | + \frac{\pi}{2} (a - b) \right)} | \overline{z} |^{a - b},
\end{multline}
which determines the large-$r$ behavior of the radial solution.  Using this result, and restricting attention to real $K$ in the regime $|z|\gg1$, the radial function $u(r)$ behaves asymptotically as a free outgoing/incoming wave,
\begin{equation}
    u\left(r\right)\simeq\sin\left(Kr-\frac{l\pi}{2}+\delta_{l}\right),
\end{equation}
where the phase shift $\delta_{l}$ encodes the effect of the potential on the partial wave and is given by
\begin{equation}
    \delta_{l}=\frac{\pi}{2}\left(l+\frac{1}{2}-\gamma_{l}\right)+\arg\Gamma\left(\frac{1}{2}+\gamma_{l}+i\eta\right).
\end{equation}

In the context of scattering phenomena involving spherically symmetric potentials, the scattering amplitude can be expanded as a series of partial waves,
\begin{equation}
    f\left(\theta\right)=\sum_{l=0}^{\infty}\left(2l+1\right)f_{l}P_{l}\left(\cos\theta\right),
\end{equation}
where the partial amplitudes $f_{l}$ are determined by the corresponding phase shifts.  Equivalently, the $S$-matrix for each partial wave is related to the phase shift by
\begin{equation}
    S_{l}=e^{2i\delta_{l}}=e^{i\pi\left(l+\frac{1}{2}-\gamma_{l}\right)}\frac{\Gamma\left(\frac{1}{2}+\gamma_{l}+i\eta\right)}{\Gamma\left(\frac{1}{2}+\gamma_{l}-i\eta\right)}.
\end{equation}

Bound states correspond to poles of the $S$-matrix.  The pole condition is therefore
\begin{equation}
    \frac{1}{2}+\gamma_{l}+i\eta=-N, \quad N=0,1,2,\dots,
\end{equation}
which yields the quantization condition for the discrete spectrum.  Introducing the auxiliary parameter
\begin{equation}
    \chi=\frac{1}{4}+\frac{l_{\alpha}\left(l_{\alpha}+1\right)}{\beta^{2}}+\delta_{r}\left(\delta_{r}+1\right),
\end{equation}
the energy eigenvalue equation can be written compactly as
\begin{multline}
   F\left(\varepsilon\right)=\left(\chi+\frac{\gamma_{s}^{2}-f^{2}e^{2}\gamma_{t}^{2}-f^{2}\delta_{t}^{2}}{g^{2}}\right)^{\frac{1}{2}} \\
   -\frac{i\left(f^{2}e\varepsilon\gamma_{t}-M\gamma_{s}\right)}{g\left(f^{2}\varepsilon^{2}-M^{2}\right)^{\frac{1}{2}}}+\left(N+\frac{1}{2}\right)=0
\end{multline}

Solving for $F(\varepsilon)=0$, we obtain the discrete energies of the bound states, corresponding to the normalizable states.

Regarding conditions for a true bound state, one must notice that, for the effective potential to have a well structure, one must have $\alpha_1(\varepsilon)<0$ and $\mathrm{min}(V^2_\mathrm{eff})<K^2(\varepsilon)<0$, imposing an interval for the found bound state energies, dependent on the Rainbow functions $f(x)$ and $g(x)$. For clarity, in the upcoming sections, we analyze each proposed case for these functions and show the procedure for obtaining results regarding energies of bound states and parameters.

\section{Case I: $f=g=\frac{1}{1-x}$}\label{Case_I}

Here we analyze specific choices for the Rainbow functions \( f(x) \) and \( g(x) \). In particular, we first consider the case in which both functions coincide and are defined as  
\begin{equation}\label{RainbowI}
    f(x)=g(x)=\frac{1}{1-x}.
\end{equation}
This choice induces a deformation of the dispersion relation and directly affects the bound-state condition. By substituting \autoref{RainbowI} into the general spectral equation, we obtain the explicit expression governing the energy eigenvalues,
\begin{multline}\label{F_Fist_case}
F\left(\varepsilon\right)=\left(\chi+\gamma_{s}^{2}\left[1-\xi\frac{\varepsilon}{\varepsilon_{p}}\right]^{2}-e^{2}\gamma_{t}^{2}-\delta_{t}^{2}\right)^{\frac{1}{2}} \\
+\frac{i\left(M\gamma_{s}\left[1-\xi\frac{\varepsilon}{\varepsilon_{p}}\right]^{2}-e\varepsilon\gamma_{t}\right)}{\left(\varepsilon^{2}-M^{2}\left[1-\xi\frac{\varepsilon}{\varepsilon_{p}}\right]^{2}\right)^{\frac{1}{2}}}+\left(N+\frac{1}{2}\right)=0.
\end{multline}
This relation implicitly determines the allowed energy levels, incorporating the effects of the Rainbow deformation together with the scalar and vector couplings and the geometric parameters encoded in the model.

\subsection{Numerical treatment}
Given the difficulty of solving \autoref{F_Fist_case} exactly, we can initially consider a numerical procedure for describing positive bound states. 

\begin{figure}[h!]
  \centering
  \subfloat[Condition function (root-finding).]{%
    \includegraphics[width=0.99\linewidth]{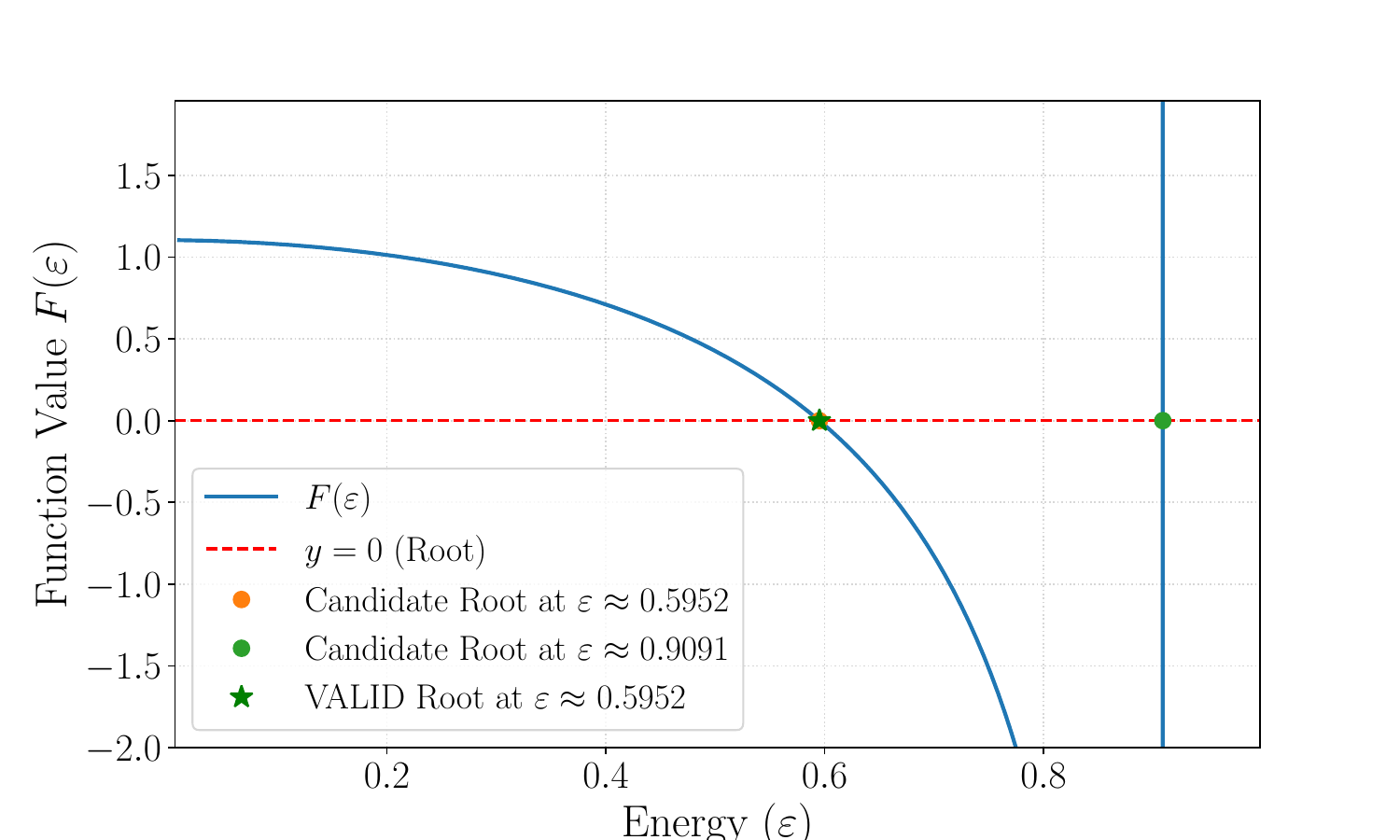}%
    \label{fig:Froot-Case1}%
  }

  \vspace{0.5cm}

  \subfloat[Effective potential for first case.]{%
    \includegraphics[width=\linewidth]{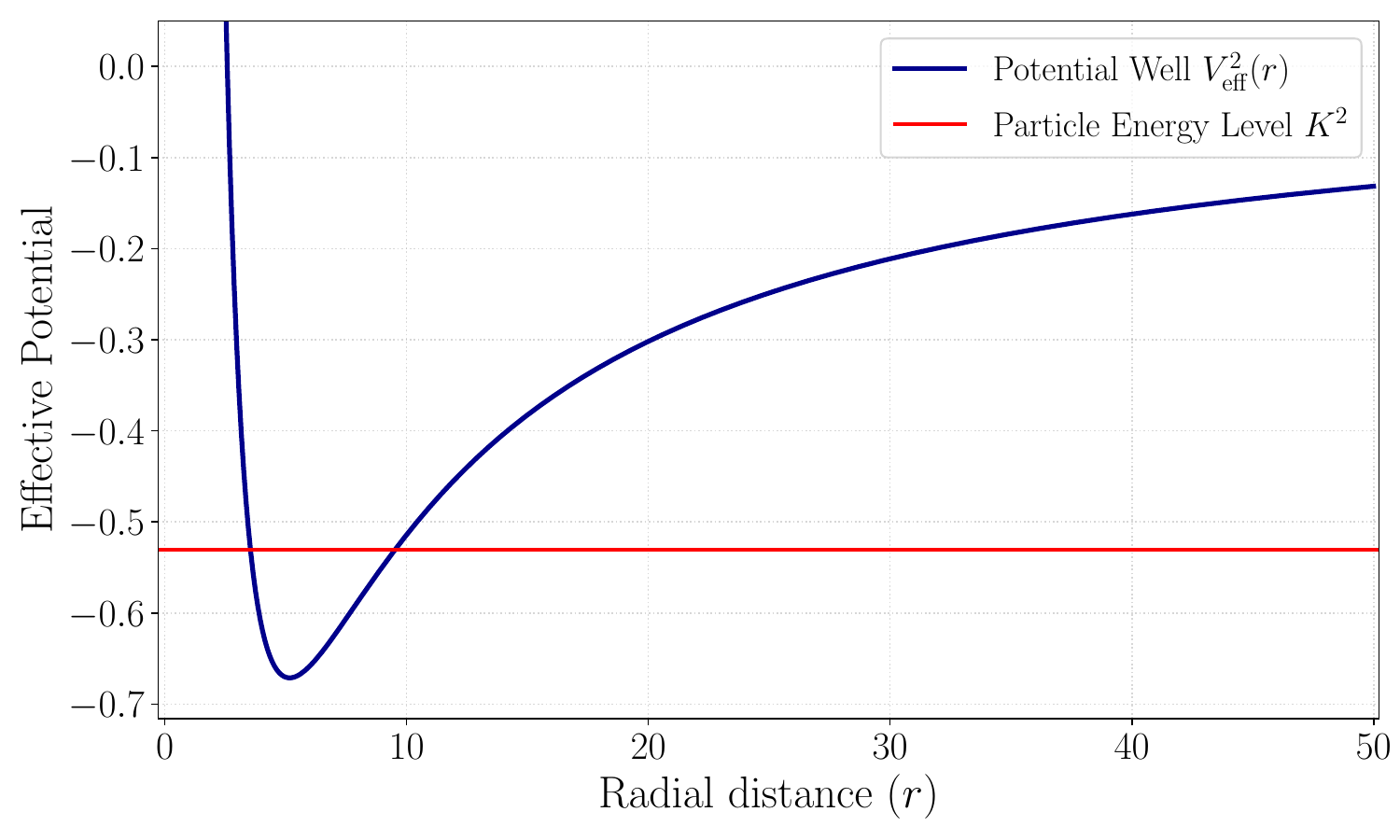}%
    \label{fig:EffPot-Case1}%
  }
  \caption{The condition function for the bound-state energy solutions and the corresponding effective potential for the first choice of Rainbow functions. The quantum numbers considered are $N=0$, $l=1$, $m=1$, and $\xi/\varepsilon_p=0.1$, which corresponds to $\alpha_1 \approx - 6.93$.}
  \label{fig:Case1-RootAndEffPotential}
\end{figure}

To ensure we obtain the proper energies for this end, we make the plot of the function $F(\varepsilon)$ in \autoref{fig:Froot-Case1} to verify possible solutions, and at the same time make sure the bound state conditions are satisfied. Moreover, after taking the proper physical solution, we also make the plot of the effective potential $V_{\text{eff}}^2$ for such a solution, given in \autoref{fig:EffPot-Case1}, where we can see the potential well and corresponding $K^2$ value relative to it. For the numerical analysis we used $\alpha = 0.8$ and $\beta = 0.8$ for the topological defects parameters, $\gamma_s = -3.85$, $\gamma_t = 0.1$, $\delta_r = 0.5$ and $\delta_t = 0.6$ for the generalized Coulomb potential parameters, $M = 1.0$ and $e = 1.0$ for the particle's parameters. 

\begin{figure}[h!]
  \centering
  \subfloat[Bound-state energies as a function of the Rainbow parameter.]{%
    \includegraphics[width=\linewidth]{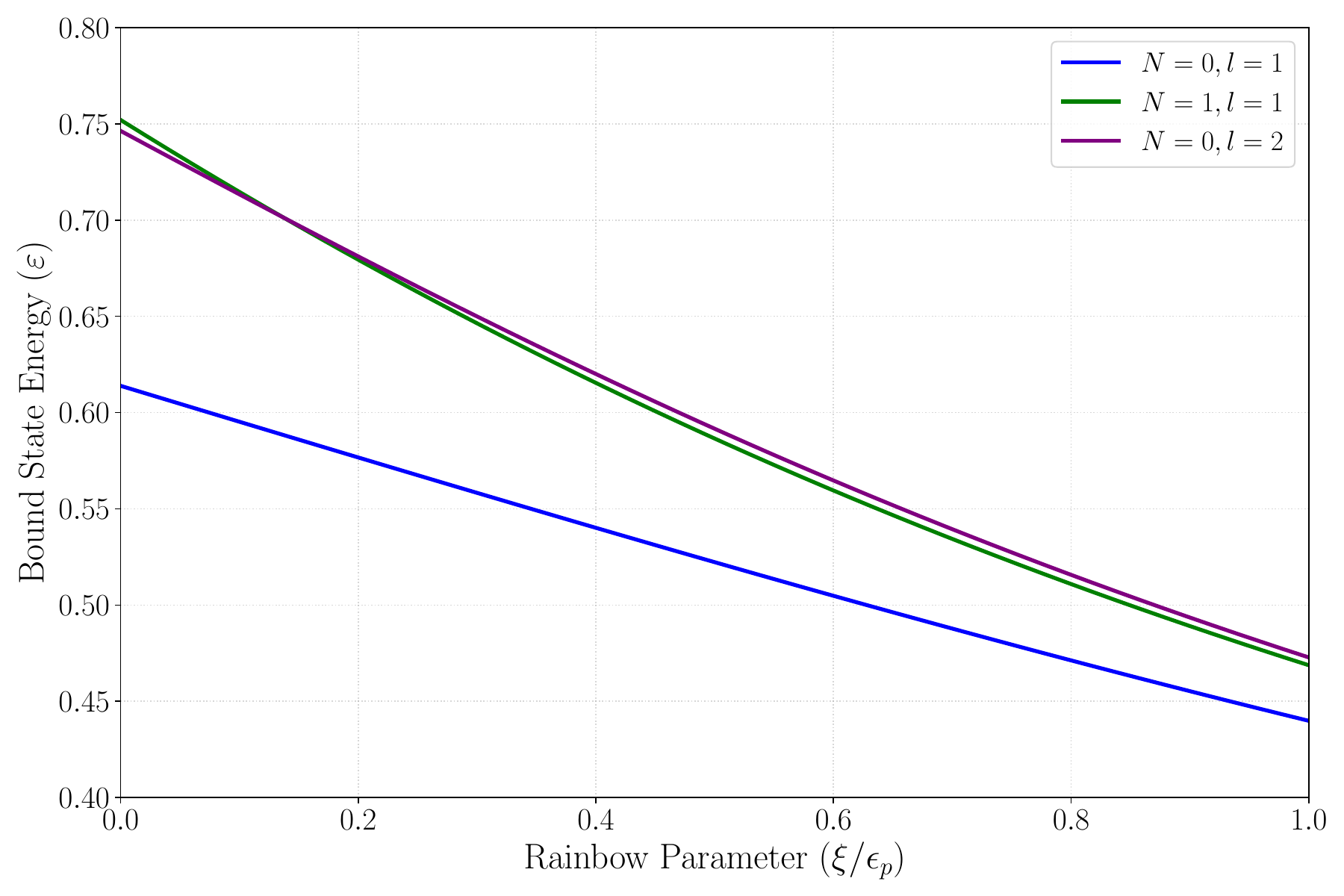}%
    \label{fig:Case1-EnergiesAgainstXi}%
  }

  \vspace{0.5cm}

  \subfloat[Energy shift induced by Rainbow gravity relative to general relativity.]{%
    \includegraphics[width=\linewidth]{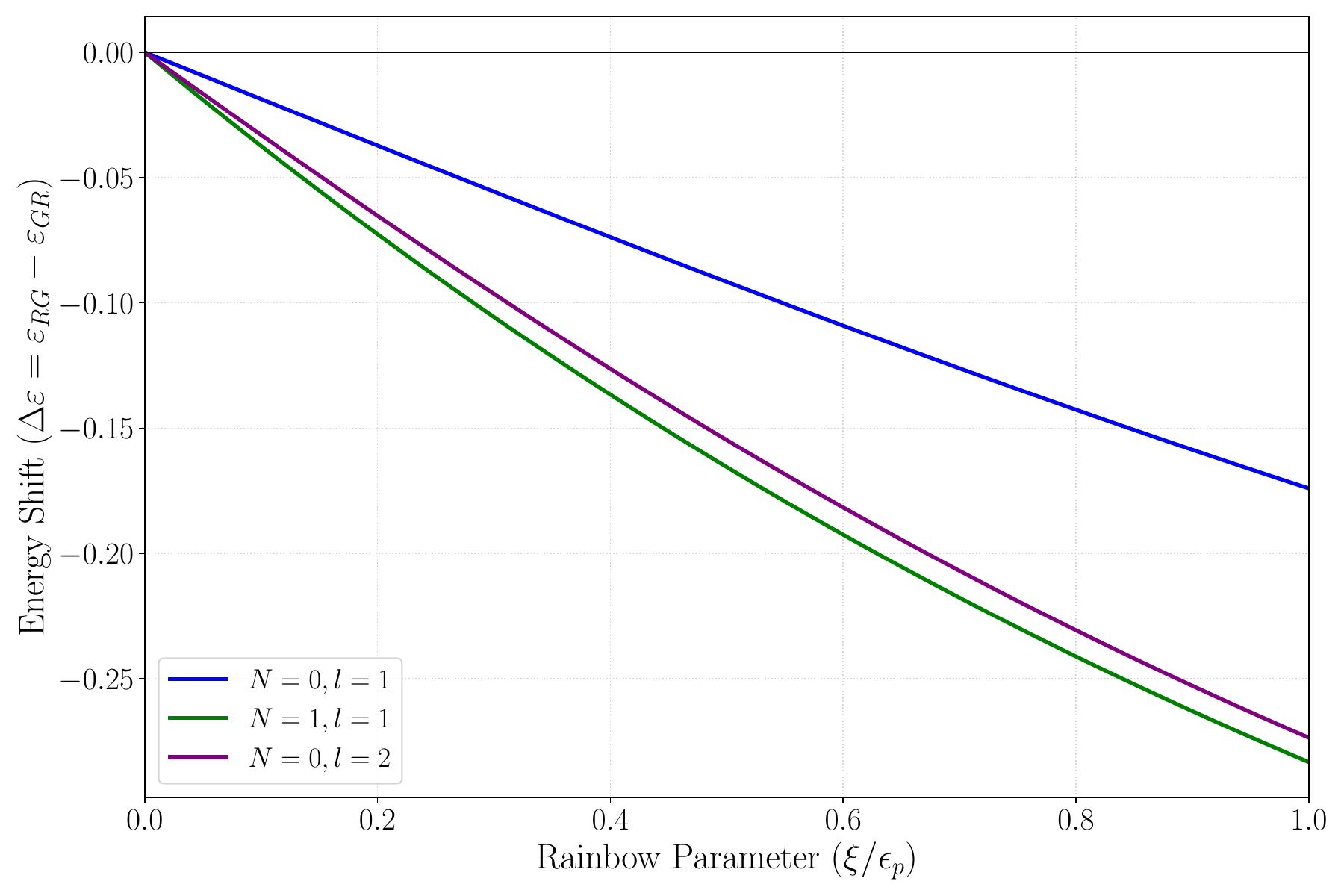}%
    \label{fig:Case1-ShiftEnergyGR}%
  }
  \caption{Bound-state energies and the corresponding energy shift induced by the Rainbow parameter for different quantum numbers, with $m=1$, in the first choice of Rainbow functions.}
  \label{fig:Case1-EnergiesAndShift}
\end{figure}

One can vary the Rainbow Gravity parameter $\xi$ and see how it affects the positive bound state energy, even comparing with the results of standard General Relativity (GR). For numerical quantities, we refer to \autoref{tab:energy_levels-Case1}, where we can see that, for the parameters considered, bound state energies could be found. Additionally, we can see how such energies vary with the $\xi$ parameter in \autoref{fig:Case1-EnergiesAgainstXi}, and compare the energy shift from GR in \autoref{fig:Case1-ShiftEnergyGR}, from which we can see that the energy in the Rainbow case is lower than the total energy in standard GR, from which we can interpret that the effect of Rainbow gravity is to make the system more bounded, and this effect increases as the parameter $\xi$ increases.

\begin{table}[h!]
\centering
\caption{Bound-state energy eigenvalues $\varepsilon$ for different quantum numbers and values of the Rainbow parameter $\xi/\varepsilon_p$, corresponding to the first choice of Rainbow functions.}
\label{tab:energy_levels-Case1}
\begin{tabular}{ccccc}
\toprule[1.5pt]
$N$ & $l$ & \multicolumn{3}{c}{$\xi/\varepsilon_p$} \\
% \cline{3-5}
    &     & 0.1 & 0.2 & 0.6 \\
\midrule[1.5pt]
0 & 1 & 0.5952 & 0.5767 & 0.5048 \\
1 & 1 & 0.7146 & 0.6794 & 0.5595 \\
2 & 1 & 0.7749 & 0.7292 & 0.5831 \\
\hline
0 & 2 & 0.7134 & 0.6811 & 0.5647 \\
1 & 2 & 0.7743 & 0.7304 & 0.5858 \\
2 & 2 & 0.8101 & 0.7586 & 0.5973 \\
\bottomrule[1.5pt]
\end{tabular}
\end{table}

Overall, the numerical analysis confirms the existence of positive bound states for the first choice of Rainbow functions and highlights the role of the Rainbow parameter in shaping the energy spectrum. The combined inspection of the condition function, the effective potential, and the energy eigenvalues shows that Rainbow gravity systematically lowers the bound-state energies. 

\subsection{Analytical treatment}
In addition to numerical analysis, we are also interested in analytical solutions for the energy spectrum $F(\varepsilon)$, so we consider the case where $\gamma_{s}/\varepsilon_{p}\approx0$, which implies
\begin{multline}
    \left(e^{2}\gamma_{t}^{2}+\mu_{N}^{2}-\frac{\mu_{N}^{2}M^{2}\xi^{2}}{\varepsilon_{p}^{2}}\right)\varepsilon^{2} \\
   -\left(2Me\gamma_{s}\gamma_{t}-\frac{2\mu_{N}^{2}M^{2}\xi}{\varepsilon_{p}}\right)\varepsilon\\
   +\left(\gamma_{s}^{2}-\mu_{N}^{2}\right)M^{2}=0 \quad,
\end{multline}
in which we define the following auxiliary parameter
\begin{equation}
    \mu_{N}=\left(N+\frac{1}{2}\right)+\left(\chi+\gamma_{s}^{2}-e^{2}\gamma_{t}^{2}-\delta_{t}^{2}\right)^{\frac{1}{2}}\quad,
\end{equation}
whose solution is given by
\begin{equation}
\begin{split}
\varepsilon_{\pm} &= \frac{M}{\left(1+\dfrac{e^{2}\gamma_{t}^{2}}{\mu_{N}^{2}}-M^{2}\dfrac{\xi^{2}}{\varepsilon_{p}^{2}}\right)}
\Bigg(\dfrac{\gamma_{s}}{\mu_{N}}\dfrac{e\gamma_{t}}{\mu_{N}}-M\dfrac{\xi}{\varepsilon_{p}} \\
&\qquad \pm\sqrt{1+\dfrac{e^{2}\gamma_{t}^{2}}{\mu_{N}^{2}}-\dfrac{\gamma_{s}^{2}}{\mu_{N}^{2}}-\dfrac{2Me\gamma_{s}\gamma_{t}}{\mu_{N}^{2}}\dfrac{\xi}{\varepsilon_{p}}+\dfrac{M^{2}\gamma_{s}^{2}}{\mu_{N}^{2}}\dfrac{\xi^{2}}{\varepsilon_{p}^{2}}}\Bigg).
\end{split}
\end{equation}

For $\xi=0$, i.e., in the absence of Rainbow gravity, we recover the known result previously obtained and analyzed in Ref.~\cite{Barbosa:2025hva}:
\begin{equation}
    \varepsilon_{\pm}=\frac{M}{\left(1+\frac{e^{2}\gamma_{t}^{2}}{\mu_{N}^{2}}\right)}\left(\frac{\gamma_{s}}{\mu_{N}}\frac{e\gamma_{t}}{\mu_{N}}\pm\sqrt{1+\frac{e^{2}\gamma_{t}^{2}}{\mu_{N}^{2}}-\frac{\gamma_{s}^{2}}{\mu_{N}^{2}}}\right).
\end{equation}

\begin{figure}[h!]
    \centering
    \includegraphics[width=\linewidth]{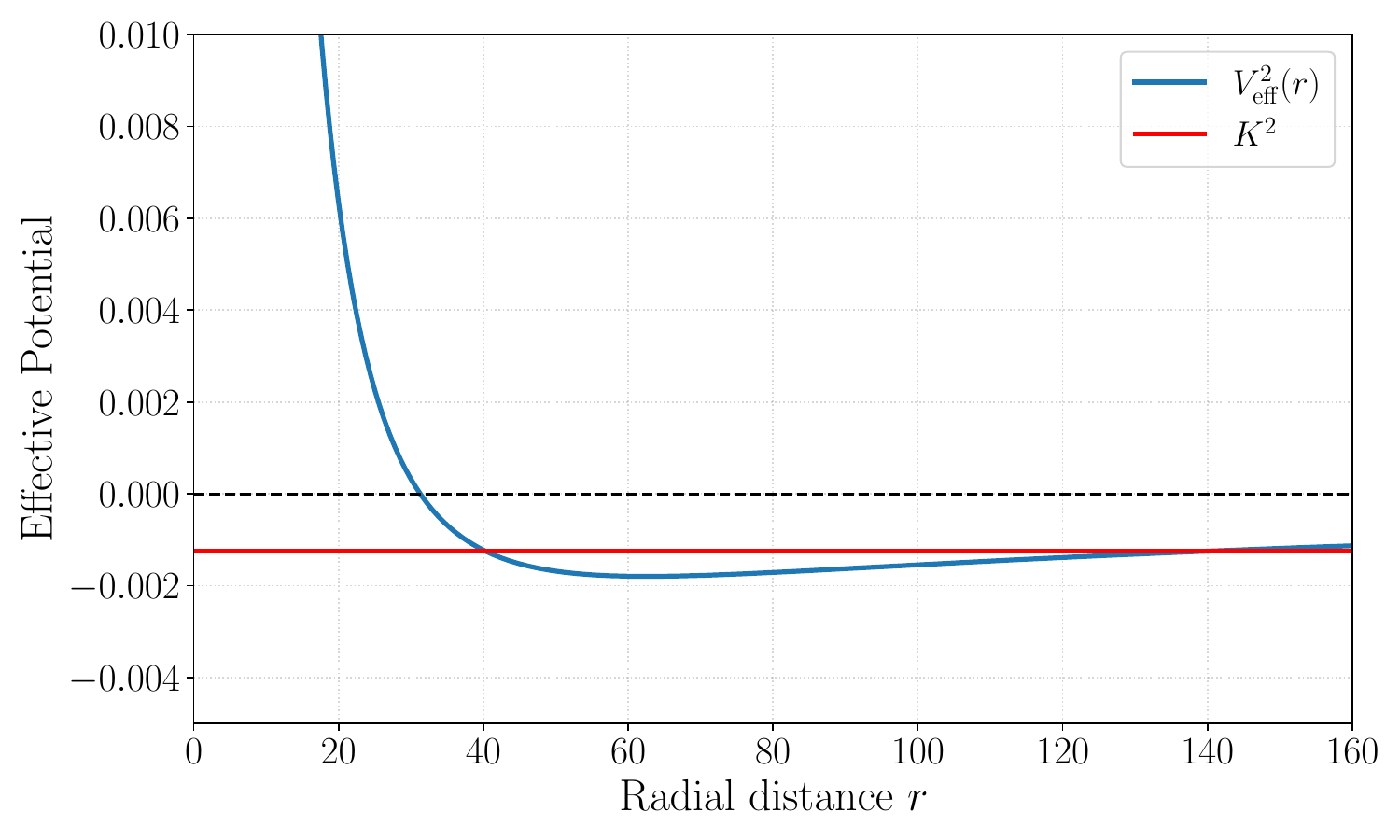}
    \caption{Effective potential for analytical solution of first case of proposed Rainbow functions for quantum numbers $N=0$, $l=2$, $m=1$ and $\xi/\varepsilon_p=0.34$.}
    \label{fig:Case1-Analytical-EffPot}
\end{figure}

\begin{figure}[h!]
    \centering
    \includegraphics[width=\linewidth]{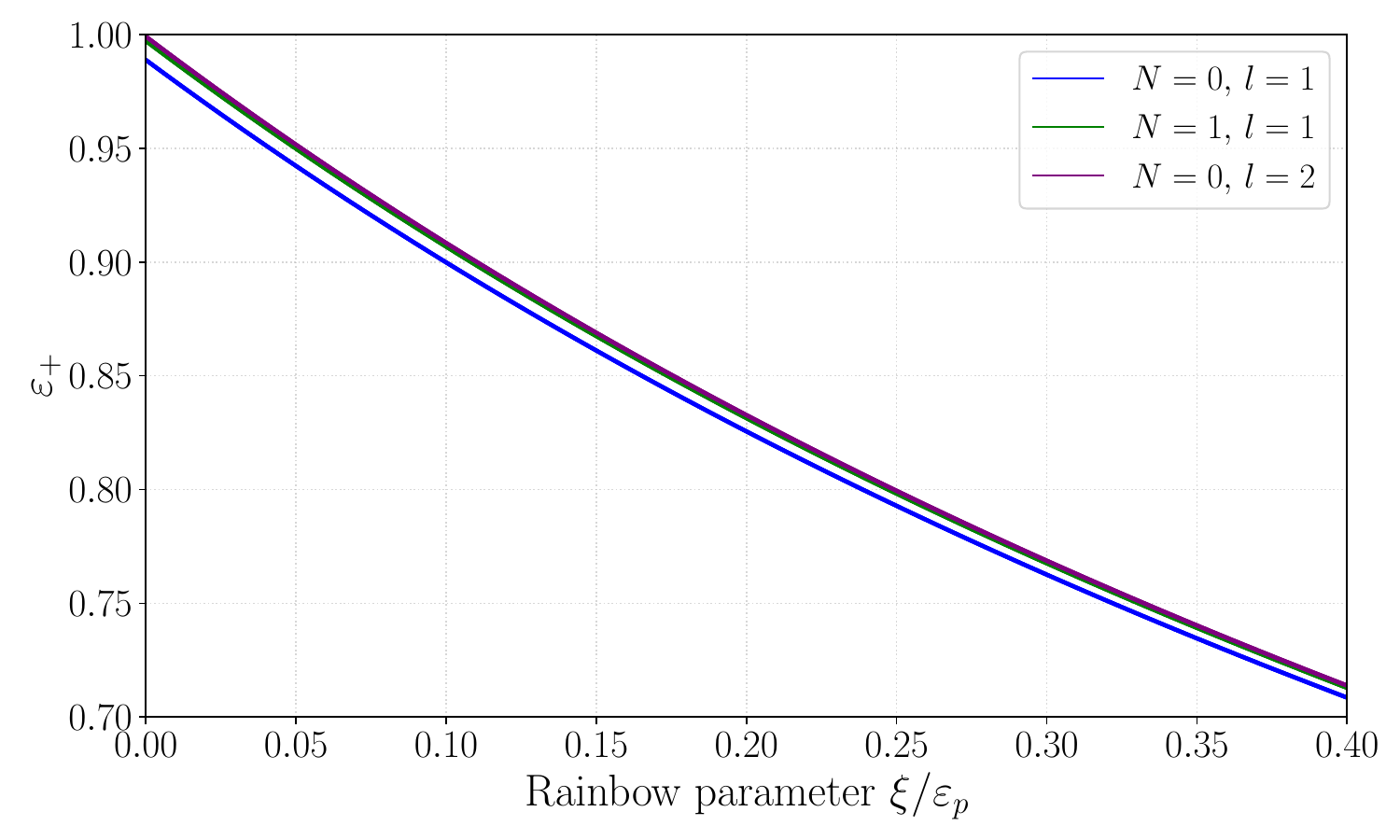}
    \caption{Bound state energies against Rainbow parameters for the first case of proposed Rainbow functions for three different sets of $N$ and $l$ quantum numbers and $m=1$.}
    \label{fig:Case1-Analytical-EnergyAgainstXi}
\end{figure}

In the regime of very low $\gamma_s$, the set of parameters that provide a valid bound state energy is more restricted. One can verify that, for the same topological defect and particle's parameters, and for the interval $0.01<\xi/\varepsilon_p<1$, a set that satisfies bound state conditions is given by $\gamma_s$ = -0.001; $\gamma_t$ = 0.15, $\delta_r$ = -0.5 and $\delta_t = 2.03$, and quantum numbers $N=\{0,1\}$, $l=\{0,1,2\}$ and $m=1$. An example of effective potential is displayed on \autoref{fig:Case1-Analytical-EffPot}, from where we can see a shallower potential well. Moreover, the behavior of bound state energies against varying $\xi$ can be seen on \autoref{fig:Case1-Analytical-EnergyAgainstXi}, from where we can see that for this new scenario closer quantum numbers have very similar energy values across all the Rainbow parameters range, in contrast with \autoref{fig:Case1-EnergiesAgainstXi}, which is due to the different effective potential shape, with minima close to zero. 

\begin{figure}[h!]
    \centering
    \includegraphics[width=\linewidth]{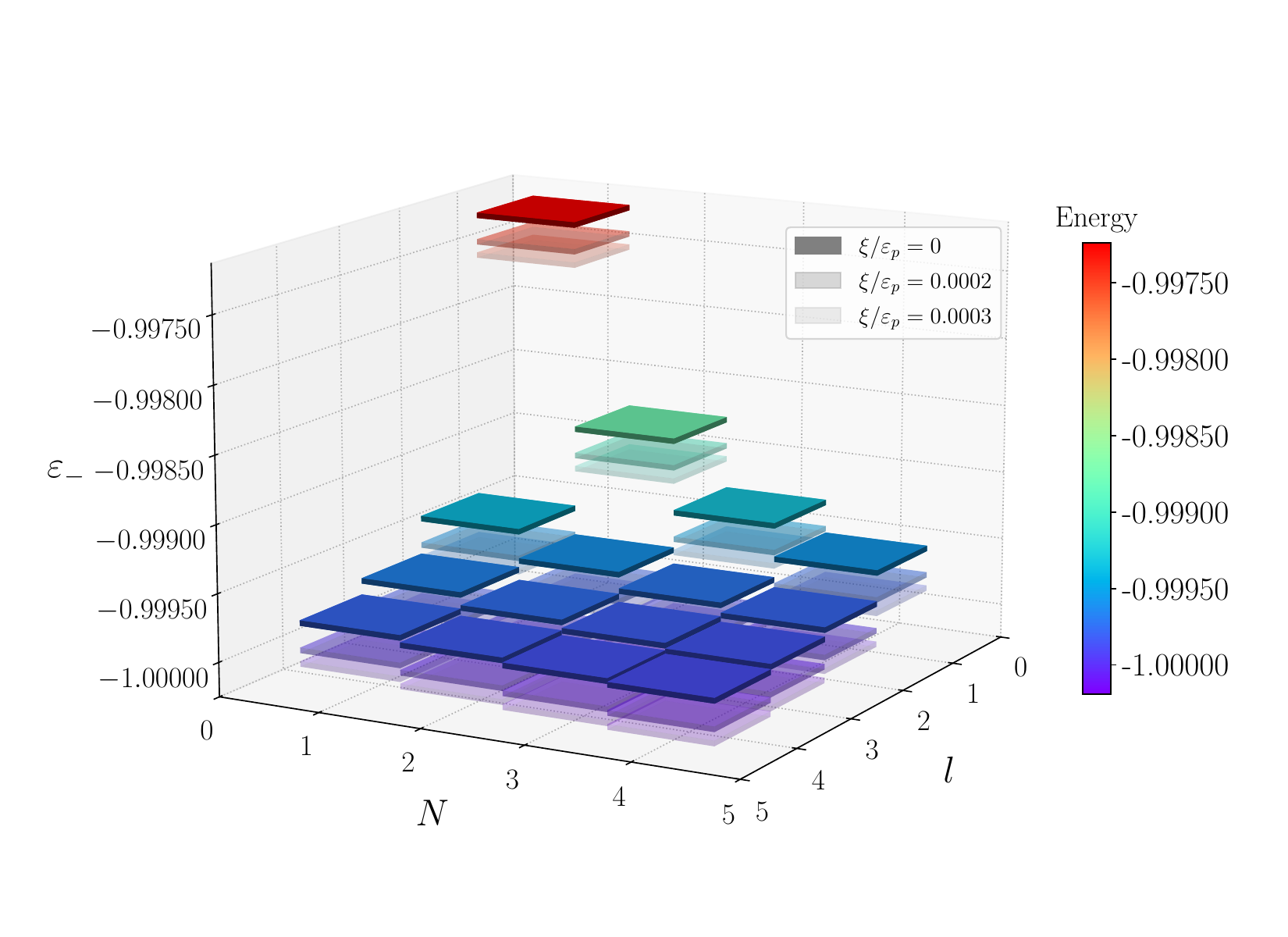}
    \caption{Negative energy spectrum dependent on quantum numbers $N$ and $l$, which take values $\{1,2,3,4\}$, for the case of pure General Relativity (solid tiles) and for different Rainbow parameters $\xi$ (semi-transparent tiles), for the second case of proposed Rainbow functions.}
    \label{fig:plot3d-case1}
\end{figure}

For the same parameters, but not particularly interested in bound states, one can observe the effect of $\xi$ on different anti-particle energy levels on \autoref{fig:plot3d-case1}, similarly done in \cite{Barbosa:2025hva,Barbosa:2025jdz}. From such a figure, we can see that just as the positive energies, the negative branch $\varepsilon_{-}$ decreases monotonically as the rainbow parameter $\xi/\varepsilon_{p}$ increases. This joint downward trend arises from two structural features of the spectrum: (i) the denominator
\begin{equation}
    D = 1 + \frac{e^{2}\gamma_{t}^{2}}{\mu_{N}^{2}} - M^{2}\frac{\xi^{2}}{\varepsilon_{p}^{2}},
\end{equation}
which is reduced as $\xi$ grows, thereby suppressing both branches simultaneously, and (ii) the linear contribution $-M \xi/\varepsilon_{p}$, stemming from the proposed Rainbow functions, inside the parentheses, which further biases both branches toward smaller values. The combined effect is that the two solutions are not symmetric under the variation of $\xi$, and no crossing behavior between $\varepsilon_{+}$ and $\varepsilon_{-}$ is observed.

\section{Case II: \( f(x)=1,\; g(x)=\sqrt{1-x^{2}} \)}\label{Case_II}

We now analyze the second choice of Rainbow functions, defined as
\begin{equation}\label{RainbowII}
    f(x)=1, \qquad g(x)=\sqrt{1-x^{2}} .
\end{equation}
For this configuration, the deformation of the background geometry is entirely governed by the function \(g(x)\), which leads to a modified spectral condition. Substituting \autoref{RainbowII} into the general energy equation, we obtain the explicit form that characterizes the energy spectrum,
\begin{multline}
    F\left(\varepsilon\right)
    =\left(\chi+\left[\gamma_{s}^{2}-e^{2}\gamma_{t}^{2}-\delta_{t}^{2}\right]
    \left(1-\xi^{2}\frac{\varepsilon^{2}}{\varepsilon_{p}^{2}}\right)^{-1}\right)^{\frac{1}{2}} \\
    +\frac{i\left(M\gamma_{s}-e\varepsilon\gamma_{t}\right)}
    {\left[\left(\varepsilon^{2}-M^{2}\right)
    \left(1-\xi^{2}\frac{\varepsilon^{2}}{\varepsilon_{p}^{2}}\right)\right]^{\frac{1}{2}}}
    +\left(N+\frac{1}{2}\right)=0 .
\end{multline}
This equation implicitly determines the allowed energy levels and serves as the starting point for the numerical analysis of both bound and scattering states in this second Rainbow scenario.

\subsection{Numerical treatment}
\begin{figure}[h!]
  \centering
  \subfloat[Condition function (root-finding).]{%
    \includegraphics[width=0.98\linewidth]{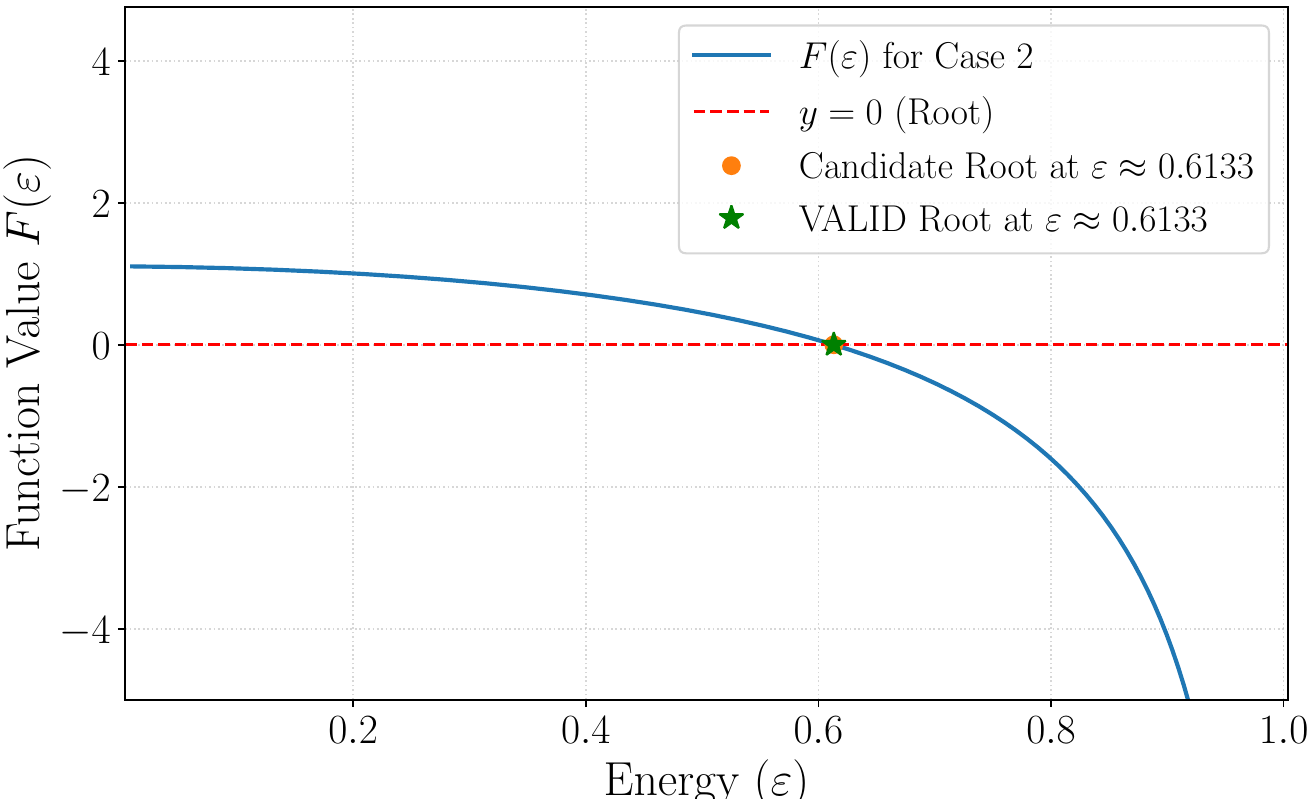}%
    \label{fig:Froot-Case2}%
  }

  \vspace{0.5cm}

  \subfloat[Effective potential for the second case.]{%
    \includegraphics[width=\linewidth]{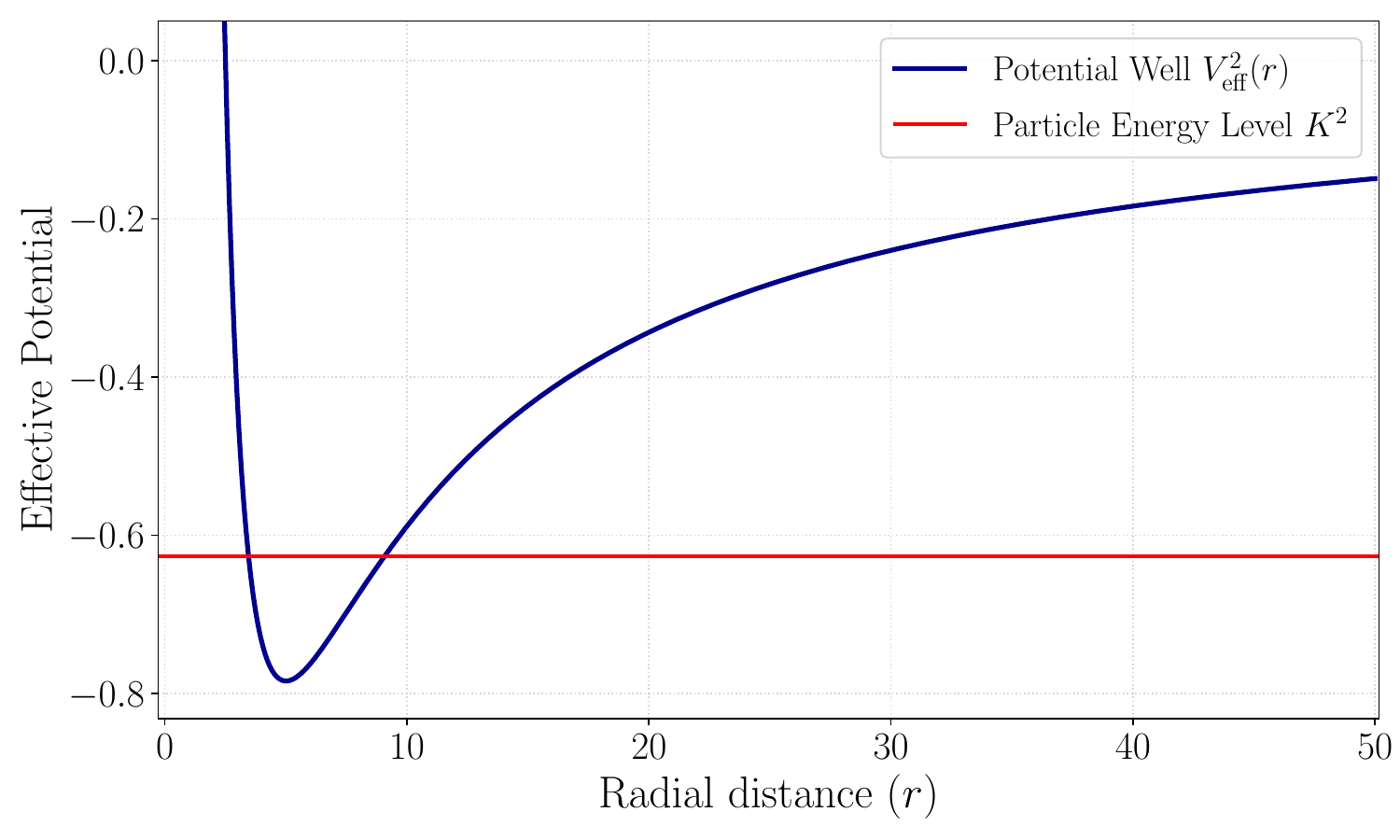}%
    \label{fig:EffPot-Case2}%
  }
  \caption{The condition function for the bound-state energy solutions and the corresponding effective potential for the second choice of Rainbow functions. The quantum numbers considered are $N=0$, $l=1$, $m=1$, and $\xi/\varepsilon_p=0.1$, with $\alpha_1\approx-7.85$.}
  \label{fig:Case2-RootAndEffPotential}
\end{figure}

This system of equations can be solved numerically using the same parameters as before. Figs. \ref{fig:Froot-Case2} and \ref{fig:EffPot-Case2} show the function $F(\varepsilon)$ and the effective potential $V_{\text{eff}}^2$ respectively, demonstrating the success of the numerical solution and confirming the existence of a positive bound state.

\begin{figure}[h!]
  \centering
  \subfloat[Bound-state energies as a function of the Rainbow parameter.]{%
    \includegraphics[width=\linewidth]{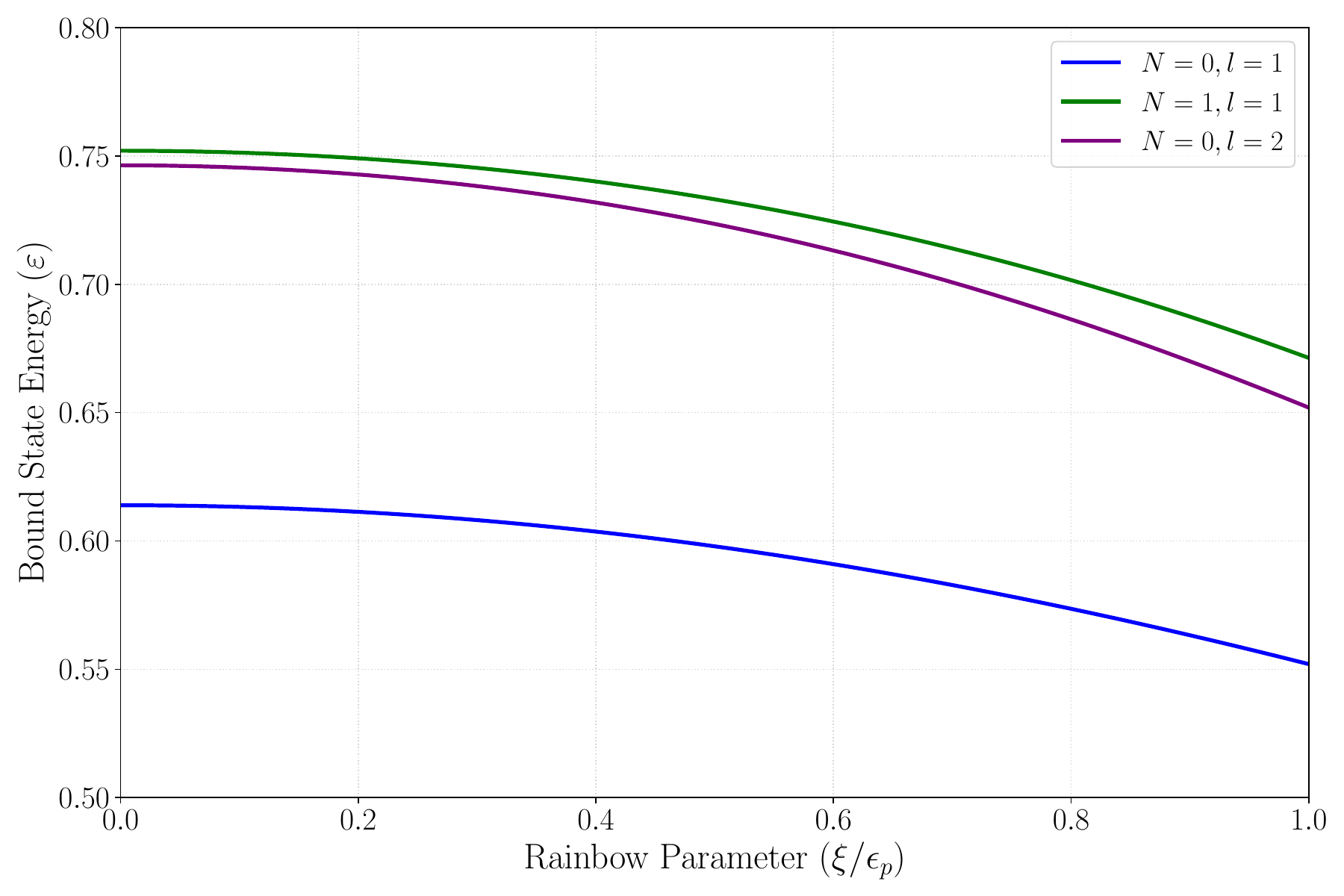}%
    \label{fig:Case2-EnergiesAgainstXi} 
    }
  \vspace{0.5cm}
  \subfloat[Energy shift induced by Rainbow gravity relative to general relativity.]{%
    \includegraphics[width=\linewidth]{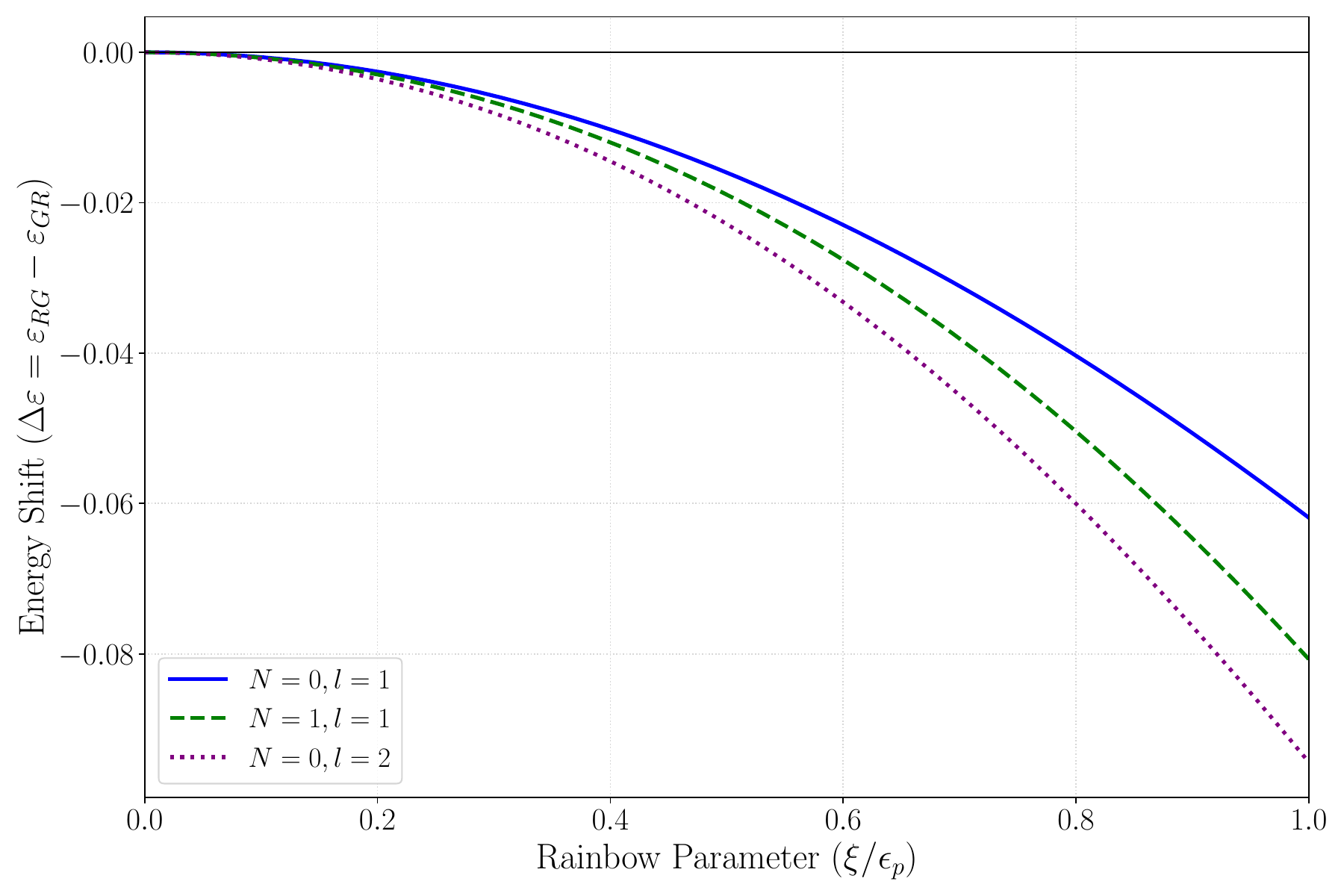}%
    \label{fig:Case2-ShiftEnergyGR}
  }
  \caption{Bound-state energies and the corresponding energy shift induced by the Rainbow parameter for different quantum numbers, with $m=1$, in the second choice of Rainbow functions.}
  \label{fig:Case2-EnergiesAndShift}
\end{figure}

Just as for the first case, we refer to \autoref{tab:energy_levels-Case2}, where we can see bound state energies found, which differ slightly from the first case. As before, we can see how such energies vary with the $\xi$ parameter in \autoref{fig:Case2-EnergiesAgainstXi}, and compare the energy shift from GR in \autoref{fig:Case2-ShiftEnergyGR}, from which we can see again, that the energy in the Rainbow case is lower than the total energy in standard GR, still making the system more bound, but this effect is lower when compared to the first case of proposed Rainbow functions.

% \begin{table}[h!]
% \centering
% \caption{Bound-state energy eigenvalues $\varepsilon$ for different quantum numbers and values of the Rainbow parameter $\xi/\varepsilon_p$, corresponding to the second choice of Rainbow functions.}
% \label{tab:energy_levels-Case2}
% \begin{tabular}{cc|ccc}
% \hline\hline
% $N$ & $l$ & \multicolumn{3}{c}{$\xi/\varepsilon_p$} \\
% \cline{3-5}
%     &     & 0.1 & 0.2 & 0.6 \\
% \hline
% 0 & 1 & 0.6133 & 0.6113 & 0.5910 \\
% 1 & 1 & 0.7513 & 0.7491 & 0.7244 \\
% 2 & 1 & 0.8241 & 0.8220 & 0.7972 \\
% \hline
% 0 & 2 & 0.7455 & 0.7428 & 0.7131 \\
% 1 & 2 & 0.8207 & 0.8183 & 0.7896 \\
% 2 & 2 & 0.8662 & 0.8640 & 0.8377 \\
% \hline\hline
% \end{tabular}
% \end{table}

\begin{table}[h!]
\centering
\caption{Bound-state energy eigenvalues $\varepsilon$ for different quantum numbers and values of the Rainbow parameter $\xi/\varepsilon_p$, corresponding to the second choice of Rainbow functions.}
\label{tab:energy_levels-Case2}
\begin{tabular}{ccccc}
\toprule[1.5pt]
$N$ & $l$ & \multicolumn{3}{c}{$\xi/\varepsilon_p$} \\
% \cline{3-5}
    &     & 0.1 & 0.2 & 0.6 \\
\midrule[1.5pt]
0 & 1 & 0.6133 & 0.6113 & 0.5910 \\
1 & 1 & 0.7513 & 0.7491 & 0.7244 \\
2 & 1 & 0.8241 & 0.8220 & 0.7972 \\
\hline
0 & 2 & 0.7455 & 0.7428 & 0.7131 \\
1 & 2 & 0.8207 & 0.8183 & 0.7896 \\
2 & 2 & 0.8662 & 0.8640 & 0.8377 \\
\bottomrule[1.5pt]
\end{tabular}
\end{table}

In summary, the numerical results for this second choice of Rainbow functions confirm the existence of positive bound states and reveal a systematic dependence of the energy spectrum on the Rainbow parameter $\xi$. Although the qualitative behavior remains similar to that observed in the first case, the corresponding energy shifts are smaller, indicating that the deformation induced by this choice of Rainbow functions leads to a weaker enhancement of the binding when compared to the previous scenario.

\subsection{Analytical treatment}
\begin{figure}[h!]
    \centering
    \includegraphics[width=\linewidth]{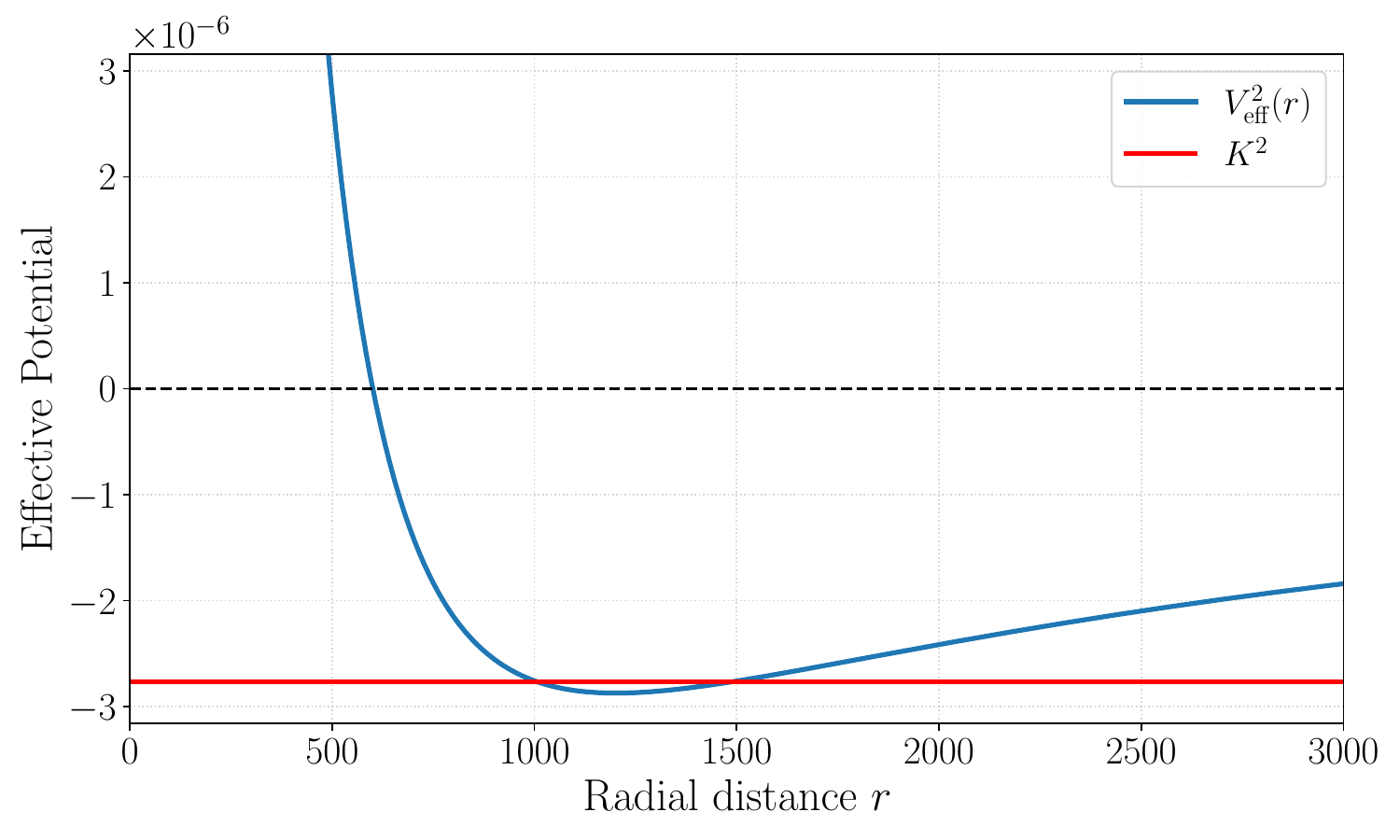}
    \caption{Effective potential for analytical solution of first case of proposed Rainbow functions for quantum numbers $N=0$, $l=1$, $m=1$ and $\xi/\varepsilon_p=0.1$.}
    \label{fig:Case2-Analytical-EffPot}
\end{figure}

\begin{figure}[h!]
    \centering
    \includegraphics[width=\linewidth]{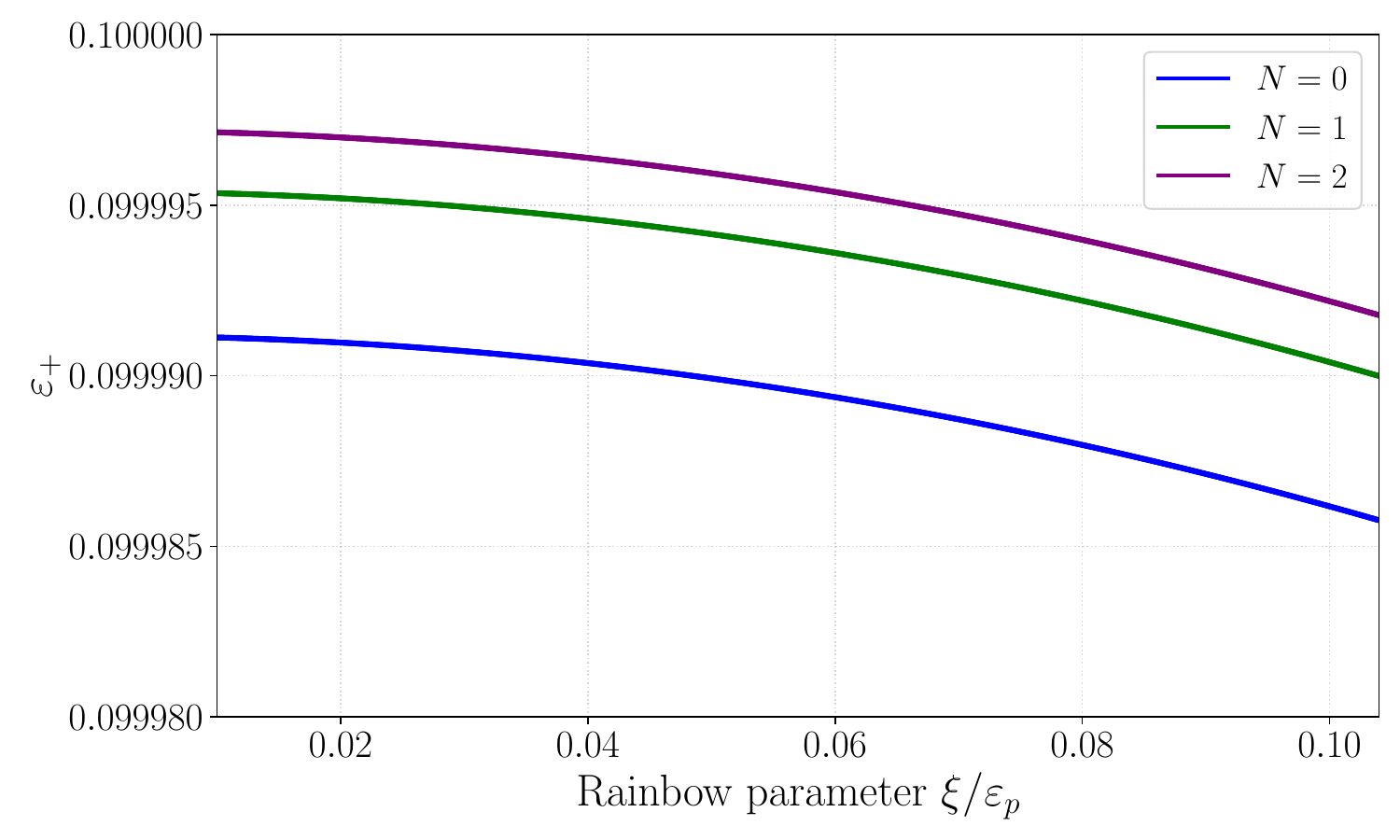}
    \caption{Bound state energies against Rainbow parameters for the first case of proposed Rainbow functions for three different sets of $N$ and $l$ quantum numbers and $m=1$.}
    \label{fig:Case2-Analytical-EnergyAgainstXi}
\end{figure}

Again, seeking analytical solutions we consider the regime $\varepsilon^{2}/\varepsilon_{p}^{2}\ll 1$. In this limit $\varepsilon^{4}/\varepsilon_{p}^{2}\ll\varepsilon^{2}$, and we further assume
$\gamma_{s}^{2}/\varepsilon_{p}^{2}\approx 0$, $e^{2}\gamma_{t}^{2}/\varepsilon_{p}^{2}\approx 0$, $\delta_{t}^{2}/\varepsilon_{p}^{2}\approx 0$. These hypotheses imply
\begin{multline}
    \left(\mu_{N}^{2}+e^{2}\gamma_{t}^{2}+\frac{\mu_{N}^{2}M^{2}\xi^{2}}{\varepsilon_{p}^{2}}\right)\varepsilon^{2} \\
    -2Me\gamma_{s}\gamma_{t}\varepsilon+\left(\gamma_{s}^{2}-\mu_{N}^{2}\right)M^{2}=0
\end{multline}
The auxiliary parameter is defined as before. The solutions of this equation are

\begin{equation}
\begin{split}
\varepsilon_{\pm} &=
\frac{M}{\left(1+\dfrac{e^{2}\gamma_{t}^{2}}{\mu_{N}^{2}}
+M^{2}\dfrac{\xi^{2}}{\varepsilon_{p}^{2}}\right)}
\Bigg(
\dfrac{\gamma_{s}}{\mu_{N}}\dfrac{e\gamma_{t}}{\mu_{N}} \\
&\qquad \pm
\sqrt{1+\dfrac{e^{2}\gamma_{t}^{2}}{\mu_{N}^{2}}
-\dfrac{\gamma_{s}^{2}}{\mu_{N}^{2}}
+\left(1-\dfrac{\gamma_{s}^{2}}{\mu_{N}^{2}}\right)
M^{2}\dfrac{\xi^{2}}{\varepsilon_{p}^{2}}}
\Bigg)
\end{split}
\end{equation}
Similarly to the previous case, for $\xi$ we recover the known case. In the regime of low $\gamma_s$ $\gamma_t$ and $\delta_t$, the set of parameters that provide a valid bound state energy is even more restricted than in the analytical configuration of the first proposed Rainbow solutions. One can verify that, for the same topological defect parameters, $M=0.1$ and $e=1$ for the particle's parameters, and for the safe interval $0.010<\xi/\varepsilon_p<0.104$, a set that satisfies bound state conditions is given by $\gamma_s = -0.01$, $\gamma_t = 0.0245$, $\delta_r$ = -0.5 and $\delta_t = 0.001$, and quantum numbers $N=\{0,1,2\}$, $l=1$ and $m=1$. An example of effective potential in this scenario is displayed on \autoref{fig:Case2-Analytical-EffPot}, from where we can see a very small difference between $K^2$ and the minimum of $V_{\mathrm{eff}}^2$, which is characteristic of the considerations herein. Additionally, the behavior of bound state energies against varying $\xi$ can be seen on \autoref{fig:Case2-Analytical-EnergyAgainstXi}, from where we can see that, similarly for the analytical considerations of the first case, closer quantum numbers have very similar energy values across the valid Rainbow parameters range, and for this analysis the bound state condition is very sensitive to the angular quantum number $l$, such that, setting it to a larger value than 1 breaks the bound state conditions.

\begin{figure}[h!]
    \centering
    \includegraphics[width=\linewidth]{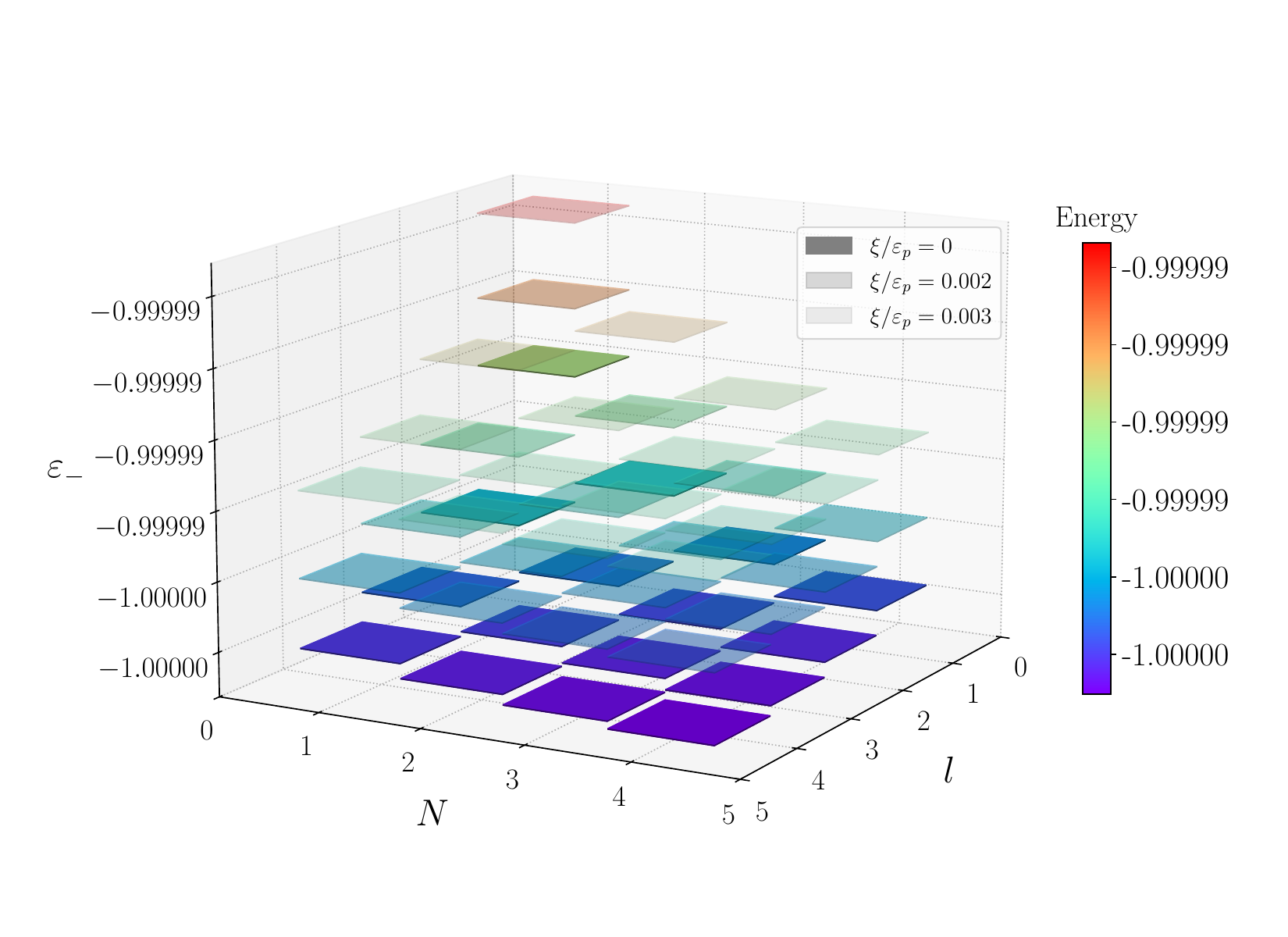}
    \caption{Negative energy spectrum dependent on quantum numbers $N$ and $l$, which take values $\{1,2,3,4\}$, for the case of pure General Relativity (solid tiles) and for different Rainbow parameters $\xi$ (semi-transparent tiles), for the second case of proposed Rainbow functions.}
    \label{fig:plot3d-case2}
\end{figure}

Similarly to what has been done for the first case, for the same parameters, one can observe the effect of $\xi$ on different anti-particle energy levels on \autoref{fig:plot3d-case2}. Similarly to the first case, the denominator now
\begin{equation}
    D = 1 + \frac{e^{2}\gamma_{t}^{2}}{\mu_{N}^{2}} + M^{2}\frac{\xi^{2}}{\varepsilon_{p}^{2}},
\end{equation}
grows with $\xi$, and the square-root structure contributes a positive quadratic term, but there is no $\xi$ proportional shifting term, and in contrast to the first case, the two branches evolve in opposite directions: the positive-energy solution $\varepsilon_{+}$ decreases, while the negative branch $\varepsilon_{-}$ is lifted toward zero as $\xi/\varepsilon_{p}$ increases.

Therefore, the qualitative distinction between Case 1 and Case 2 is rooted in the analytic structure of the rainbow functions: the specific choice of \(f(x),g(x)\) determines whether both branches shift coherently (Case 1) or split symmetrically (Case 2) under the variation of \(\xi/\varepsilon_{p}\).

\section{Discussion and conclusions}\label{Discussion_and_conclusions}

In this work, the relativistic dynamics of scalar bosons was analyzed in a spacetime containing a cosmic string and a global monopole within the framework of gravity’s rainbow. After constructing the effective metric and deriving the Klein–Gordon equation with scalar, vector, and nonminimal couplings, the problem was reduced to angular and radial sectors, allowing a detailed investigation of scattering and bound states under generalized Coulomb-type interactions. Bound states were identified through the pole structure of the S-matrix, establishing a direct connection between the spectrum and the geometric, coupling, and rainbow parameters of the model.

The effects of Rainbow gravity were explored by considering two specific choices of rainbow functions. For each case, the bound-state conditions were studied numerically by analyzing the implicit spectral equations and the associated effective potentials, which made it possible to verify the existence of bound states and to track the dependence of the energy levels on the rainbow parameter. The numerical analysis was complemented by analytical treatments in restricted regimes, where suitable approximations allowed the derivation of explicit expressions for the energy spectrum. These analytical results clarified the role of the rainbow deformation and showed how known expressions are recovered when rainbow effects are suppressed. Taken together, the numerical and analytical analyses provide a consistent description of how the combined presence of topological defects and energy-dependent spacetime deformations influences the spectral properties of scalar bosons.

\section{Acknowledgements}\label{Acknowledgements}
L.G.B. and J.V.Z. acknowledge the financial support from CAPES (process numbers 88887.968290/2024-00 and 88887.655373/2021-00, respectively). L.C.N.S. would like to thank Conselho Nacional de Desenvolvimento Científico e Tecnológico - Brazil (CNPq) for financial support under Research Project No. 443769/2024-9 and Research Fellowship No. 314815/2025-2.

\bibliographystyle{unsrturl}
\bibliography{sample}

\end{document}